\documentclass[12pt]{amsart}
\usepackage{amssymb}
\usepackage{amsmath,amsthm,amsfonts}
\usepackage[applemac]{inputenc}
\usepackage{graphicx}
\usepackage{bbm}
\usepackage{pdfsync}
\usepackage{fancyhdr}
\usepackage{mathrsfs}
\usepackage{slashed}
\usepackage{color}
\usepackage{xcolor}
\usepackage[breaklinks,colorlinks,backref]{hyperref}

\hypersetup{
colorlinks, 
linktoc=all, 
linkcolor=blue, 
citecolor=hyptxt,
urlcolor=blue}
\hypersetup{
citebordercolor=,
filebordercolor=green,
linkbordercolor=blue
}
\definecolor{hyptxt}{rgb}{0.7, 0.4, 0.9}

\usepackage{bibtopic}

\usepackage[full]{textcomp} 
\usepackage{fbb} 
\usepackage[varqu,varl]{inconsolata}
\usepackage[libertine,bigdelims,vvarbb]{newtxmath}
\usepackage[cal=boondoxo]{mathalfa}
\usepackage[T1]{fontenc}
\useosf

\newtheorem{prop}{Proposition}[section]

\newcommand{\beprop}{\begin{prop}}
\newcommand{\enprop}{\end{prop}}
\newcommand{\bprf}{\begin{proof}}
\newcommand{\eprf}{\end{proof}}

\newcommand{\ket}[1]{|\kern.3ex#1\kern.3ex\rangle}
\newcommand{\bra}[1]{\langle\kern.3ex #1 \kern.3ex|}
\newcommand{\scalar}[2]{\langle\kern.3ex #1 \kern.3ex|\kern.3ex#2\kern.3ex\rangle}

\def\R{\mathbb{R}}

\def\C{\mathbb{C}}

\def\ii{\mathrm{i}}
\def\ud{\mathrm{d}}

\def\bu{\mathbbm{1}}

\def\ud{\mathrm{d}}
\definecolor{hervecolor}{rgb}{0.8,0,0.7}

\numberwithin{equation}{section}

\def\R{{\rm I\hspace{-.15em}R}}

\def\1{\mbox{I\hspace{-.15em}1}}

\def\setC{\mathbb{C}}

\def\b{\begin{equation}}
\def\e{\end{equation}}

\usepackage[compat=1.0.0]{tikz-feynman}
\usepackage{float}

\begin{document}
\date{\today}
\title{Axiomatic de Sitter quantum Yang-Mills theory \\ with color confinement and mass gap}
\author{M.V. Takook}

\address{\emph{ APC, UMR 7164}\\
\emph{Universit\'e Paris Cit\'e} \\
\emph{75205 Paris, France}}

\email{ takook@apc.in2p3.fr}

{\abstract{The analyticity property of de Sitter's quantum Yang-Mills theory in the framework of Kerin space quantization, including quantum metric fluctuation, is demonstrated. This property completes our previous work regarding quantum Yang-Mills theory in de Sitter's ambient space formalism, and we can construct an axiomatic quantum field theory similar to Wightman's axioms. The color confinement is proven for the general case, which was previously approved in the early universe. It is shown by using the interaction between gluon fields and the conformal sector of the gravitational field, which is a massless minimally coupled scalar gauge field. The gluon mass results from the interaction between the gluon fields and the massless minimally coupled scalar field as a conformal sector of the gravitational field and then the symmetry-breaking setting due to the vacuum expectation value of the scalar field.}}

\maketitle

{\it Proposed PACS numbers}: 04.62.+v, 98.80.Cq, 12.10.Dm



\section{Introduction}
\label{intro}

A challenging problem of QFT is constructing an axiomatic quantum Yang-Mills theory, similar to Wightman's axiom approach, which includes color confinement and mass gap\footnote{It is one of the seven Millennium Prize Problems in mathematics that the Clay Mathematics Institute stated in May $24, 2000$.} \cite{jawi}. From the Wilson loops in lattice gauge theory, which explains the confinement phase, one could conclude that the confinement problem relates to the spacetime geometry or gravitational field. That means the source of color confinement may come from the geometry of spacetime. In this regard, the QCD in curved spacetime has been considered by many authors over the years. Finally, the precise construction of the color confinement problem was solved in the one-loop approximation by Kharzeev et al., see \cite{khletu} and the references therein.

Kharzeev et al. constructed an effective low-energy Lagrangian of gluodynamics on a curved conformal background that satisfies all constraints imposed by the Renormalization Group. Their model is the scale and conformally invariant in the vanishing vacuum energy density limit. It matches the perturbative theory at short distances. Color fields are dynamically confined, and the strong coupling freezes at distances more significant than the glueball size. The main idea of their model is the coupling between a scalar field (the conformal sector of the metric) with the massless vector fields (gluon fields), which mediate the strong interaction. Nevertheless, their model is only renormalizable in one-loop approximation and cannot be formulated in an axiomatic Wightman-like QFT framework.

The existence of a covariant and renormalizable quantum Yang-Mills theory within the framework of the Krein space quantization (including quantum metric fluctuation) has recently been suggested in de Sitter (dS) ambient space formalism \cite{taga22}. We obtained that the ghost fields are the massless minimally coupled (mmc) scalar fields. The fundamental part of our model is the covariant quantization of the mmc scalar field based on the Krein space quantization. We obtained that the source of the mass gap comes from the infrared limit of the mmc scalar field in the interaction with the gluon fields, and also, the origin of the color confinement gets out from the spacetime geometry. The confinement of colors was only discussed in the early universe when the effect of the gravitational field was so significant. Also, the analyticity property does not discuss in previous work \cite{taga22}. This property is essential for constructing the interaction QFT in an axiomatic approach similar to Wightman's axiom, which will be studied in this article.

Recently, we formulated the mmc scalar field in the dS ambient space formalism as a gauge potential or connection field \cite{ta223}. We construct the scalar gauge theory by helping an arbitrary constant five-vector field $B^\alpha$ analogous to the standard gauge theory. It is shown that the dS ambient space formalism permits us to unify the scalar and vector gauge fields. The scalar gauge field can be interpreted as the conformal sector of the gravitational field. In the Landau gauge, the conformal sector is the mmc scalar field \cite{ta09,enrota}. This field may be interpreted as the connection between the different dS hyperboloids in quantum dS geometry. It breaks the symmetry of a specific dS hyperboloid \cite{ta223}.

In this letter, by using the ideas of the three papers mentioned above: $1)$ the coupling of the conformal sector of the metric with the gluon fields \cite{khletu}, $2)$ Yang-Mills theory in Krein space quantization \cite{taga22}, and $3)$ the scalar gauge field as the conformal sector of the gravitational field \cite{ta223}; an axiomatic quantum Yang-Mills theory in the framework of the Krein space quantization (including quantum metric fluctuation) with a mass gap and color confinement is constructed in the dS ambient space formalism. This construction is renormalizable and analytic and can be applied to the current universe. It is essential to insist that the ambient space formalism and the Kerin space quantization make it possible to build a coherent model for the quantum Yang-Mills theory.

It is interesting to note that some authors have already discussed these ideas. The couplings of Lorentz vector and scalar potentials for explaining the confinement problem of spinless particles in $1+1$ dimensions have also been studied in \cite{dcas}. Recently, the dependence of color confinement on dS and Anti-dS geometry and especially on the conformal sector ({\it i.e.} scalar field potential) is also considered in \cite{kico,kipova}.


\section{Axiomatic QFT}
\label{NAGT}

The axiomatic QFT of the massive field can be constructed from the Wightman axioms in Minkowski spacetime \cite{strweit}. It can be directly generalized to the massive field in dS spacetime (principal series representation) by replacing the positive energy condition with a certain geodesic spectral condition \cite{brmo}. Nevertheless, some axioms must be changed for the massless vector field due to the gauge invariance. In Minkowski space, Gupta introduced four types of photons, and a covariant quantization for massless vector field was obtained \cite{gupta}. However, the scalar photon has negative norms. Then for getting a similar approach to Wightman's axioms, the positivity conditions must be replaced with the existence of an indefinite sesquilinear form; see \cite{stroch} and the references therein. Then this procedure of the existence of four types of photons was generalized to the dS spacetime, and an axiomatic QED in dS space was constructed \cite{gagarota}. Now we would like to generalize it to QCD. Here we only discuss the differences between QED and QCD in Minkowski and dS spaces.

We know that in Minkowskian spacetime, the four types of photons satisfy the same field equations:
$$ \Box A_\mu^{(i)}=0, \;\;\; i=1,2,3,4,$$
where $ \Box=\partial_\mu\partial^\mu$ is the Laplace-Beltrami operator on Minkowski spacetime. In this case, all modes propagate on the Minkowskian light cone. Nevertheless, in dS spacetime, the transverse photons ($K_{\alpha}^{(t)}, t=1,2$) propagate on the dS light cone, and the two others (scalar and gauge modes $\phi_s, \phi_g$) does not propagate only on the dS light cone. These modes satisfy the following equations \cite{gagarota}:
\b \label{gcsll} \left\{\begin{array}{clcr} \phi_s\equiv &\partial \cdot K \neq 0\, ,\: \;\; \Longrightarrow\;\;\;\; \;\;\;\; \;\;\;\; \;\;\;\; \Box_H \phi_{s}=0,\\
\;& \partial \cdot K=0 \,, \;\;\;\Longrightarrow \; \left\{\begin{array}{clcr} \partial_\alpha^\top \phi_g, &\; \Box_H \phi_{g}=0\,,\\
K_{\alpha}^t , & (\Box_H +2H^2)K^{(t)}= 0 \, ,
\end{array} \right.
\end{array} \right.
\, \e
where $\phi_s$ and $\phi_g$ can be associated with the scalar photon and scalar pure gauge mode, respectively \cite{gagarota}. $H$ is the Hubble constant parameter, and $ \Box_H$ is the Laplace-Beltrami operator on dS spacetime. In ambient space formalism, it can be written in the following form:
$$ \square_H=\partial^\top\cdot\partial^\top\,; \;\;\; \partial_\beta^\top =\theta_{\alpha \beta}\partial^{\alpha}=
\partial_\beta + H^2 x_\beta x\cdot\partial\,; \;\; \alpha, \;\beta\equiv 0,1, \cdots,4\, , $$
where $ \theta_{\alpha \beta}=\eta_{\alpha \beta}+
H^2x_{\alpha}x_{\beta}$ is the transverse projector on the dS hyperboloid; for ambient space notation, see \cite{taga22}. The scalar and pure gauge modes are the mmc scalar fields. They are auxiliary unphysical states. The transverse modes satisfy the massless conformally coupled scalar field equation, in which they are the physical states and propagate on the dS light cone. In the previous paper \cite{gagarota}, the quantization was done in the standard method. However, in the indecomposable representation, the scalar and the pure gauge modes are quantized in the Krein space quantization. Due to the coupling of the propagator to the conserved current, scalar and gauge photons are entirely decoupled from the theory, and they can not appear in the internal line of the Feynman diagram. An axiomatic quantization of the free mmc scalar field in the Krein space was constructed previously in \cite{gareta00}.

However, the situation for QCD is rather different, and the ghost modes appear, which are also mmc scalar fields \cite{taga22}. Although these modes can be eliminated in the external line of the Feynman diagrams by imposing the reality condition, their presence in the internal line can not be ignored, similar to the Minkowskian counterpart. In this case, contrary to QED, due to the ghost fields, the quantization must be done entirely in the Krein space to obtain a covariant quantization or an axiomatic quantum field theory. For simplicity, the scalar field is discussed, but it can be easily generalized to the other spin fields.

Using a) dS covariance, b) locality, and c) existence of an indefinite sesquilinear form, the free quantum field theory can be constructed on the Krein space (Hilbert $\oplus$ anti-Hilber spaces). We begin with the following infinite-dimensional local closed algebra:
\b \label{algebcom} [\phi(x), \phi(x')]\equiv G(x,x') \bu \, ,\e
where $G(x,x')=\mathcal{W}(x,x')-\mathcal{W}(x',x)$ is the commutation two-point function, and it is zero for space-like separate points. $ \bu$ is the identity operator on the Krein space, which is constructed on the operator algebra \eqref{algebcom}. $\mathcal{W}(x,x')=\langle \Omega \mid \phi(x)\phi(x') \mid \Omega\rangle$ is the two-point function and $ \mid \Omega\rangle$ is the Krein-Fock vacuum state \cite{gareta00}. For obtaining a well-defined and normalizable field operator, they must be defined in a distribution sense (tempered distributions) on an open set $\mathcal{O}$ of spacetime \cite{brmo}. For any test function $f(x) \in \mathcal{ D}(X_H)$, we have an indefinite sesquilinear form that is defined by
\begin{equation} \int _{X_H \times X_H}
f^*(x)\mathcal{ W}(x,x')
f(x')\ud \sigma(x)\ud \sigma(x')\, ,\end{equation}
where $ f^*$ is the complex conjugate of $f$. $X_H$ is a $4$-dimensional hyperboloid embedded in the $5$-dimensional Minkowskian spacetime with the equation:
\b \label{dSs} M_H=\{x \in \R^5| \; \; x \cdot x=\eta_{\alpha\beta} x^\alpha
x^\beta =-H^{-2}\},\;\; \alpha,\beta=0,1,2,3,4, \e
where $\eta_{\alpha\beta}=$diag$(1,-1,-1,-1,-1)$. $\mathcal{ D}(X_H)$ is the space of $C^\infty$ functions with compact support in $X_H$ and with values in $\setC $. $\ud \sigma(x)$ is dS invariant volume element. For the construction of the Hilbert space from the operator's field algebra in dS space, see \cite{tagahu}.

Then the field operator can be divided into two parts: positive norm state and negative norm state, which act on the Hilbert space and anti-Hilbert space, respectively:
$$ \phi(f)=\frac{1}{\sqrt{2}}\left[\phi_p(f)+\phi_n(f)\right]\,.$$
It is logical to write the field operator into the creation $\phi^+$ and the annihilation parts $\phi^-$:
$$ \phi_p(f) =\phi^-_p(f)+\phi^+_p(f)\, , \; \phi_n(f)=\phi^-_n(f)+\phi^+_n(f)\, . $$
$\phi^+_p(f)$ creates a positive norm state, and $\phi^-_p(f)$ annihilates a positive norm state from the Hilbert space. $\phi^+_n(f)$ creates a negative norm state, and $\phi^-_n (f)$ annihilates a negative norm state from the anti-Hilbert space.

In Krein space quantization, the two-point function is the imaginary part of the two-point function of the positive mode solutions, or standard method \cite{gareta00,ta3}:
\b \label{ktpf} \mathcal{W}(x,x')=\mathcal{W}_p(x,x')+\mathcal{W}_n(x,x')= \ii \mathrm{Im} \mathcal{W}_p(x,x')=\frac{1}{2} G(x,x'),\e
where $\mathcal{W}_n(x,x')=-\mathcal{W}_p^*(x,x')$. The two-point commutation function in the two models is the same $G(x,x')=G_p(x,x')$; therefore, free-field quantization in both models gives the same results.

To generalize the axiomatic approach to the quantization of Kerin's space to interaction quantum field theory, such as Yang-Mills theory, we need to prove the analyticity properties of the vacuum expectation value of the field operators product \cite{strweit,brmo}:
\b \label{vevt} \mathcal{W}_N(x_1,\cdots,x_N)\equiv \langle\Omega\mid \,\phi(x_1)\phi(x_2) \cdots \phi(x_N) \mid \Omega\rangle \, .\e
First, we recall a theorem regarding the analytic/holomorphic nature of sums, differences, and products of analytic functions.\newline
{\bf Analyticity theorem:} Let $A\subseteq \C$ be an open set, $k\in \C$, and let $f,g:A\longrightarrow \C$. Let $f$ and $g$ be analytic on $A$, which are differentiable at every point $z_0\in A$. Then:\newline
A) $f+g$ is analytic on $A$ and $(f+g)'(z)=f'(z)+g'(z)$ for all $z\in A$.\newline
B) $kf$ is analytic on $A$ and $(kf)'(z)=kf'(z)$ for all $z\in A$. \newline
C) $fg$ is analytic on $A$ and ($fg)'(z)=f(z)g'(z)+f'(z)g(z)$ for all $z\in A$.

The Wightman two-point function $\mathcal{W}_p(x,x')$ is analytic in the tuboid $\mathcal{T}_{12}$, to prove analytic property and define the tuboid, see \cite{brmo}. $\mathcal{W}_n(x,x')$ is also analytic but in the tuboid $\mathcal{T}_{21}$. Then in Krein space quantization, the two-point function \eqref{ktpf} is not analytic. However, it is free from all infrared and ultraviolet divergence except the singularity of the light cone. Nevertheless, we know this type of singularity can be manipulated by metric quantum fluctuation \cite{ta02}. Then in Krein space quantization which includes quantum metric fluctuation, the two-point function $\langle\mathcal{W}(x,x')\rangle$ (or equivalently $\langle G(x,x')\rangle$) is well-defined. It is differentiable on the dS hyperboloids; see equation $(2.5)$ in \cite{ta02} for Minkowskian space and equation $(5.15)$ in \cite{taga22} for dS space. The expectation value is taken over the first-order quantum metric fluctuations \cite{for,fosv,for3}. $\langle\mathcal{W}(x,x')\rangle$ is a regular function on the dS hyperboloids in Krein space quantization, {\it i.e.} non-singular and differentiable. Therefore, it is analytic on the dS hyperboloids. In this case, the time-ordered product two-point function $\langle G_T(x,x')\rangle$ is also analytic.

From the Feynman path integral on curved spacetime \cite{park} and also the Feynman-type algebra for dS spacetime \cite{brmo,brepmo2}, $G_T(x_1,\cdots,x_N)$ can be written in terms of a summation and multiplication of the two-point functions $G_T(x,x')$ in Krein space quantization. Then $G_T(x_1,\cdots,x_N)$ is only singular on the light cones, which can be
manipulated by metric quantum fluctuation. Since the two-point function $\langle G_T(x,x')\rangle$ is analytic and using the calculation of correlation function and the Analyticity theorem, the $N$-point correlation function $\langle G_T(x_1,\cdots,x_N) \rangle$, is well-defined on the dS hyperboloid and then it is analytic. For more details on calculating the correlation functions in light cone fluctuation, see \cite{fosv}.

Therefore, the axiomatic procedure of quantum interaction field theory can be completed similarly to Wightman's axioms by using ambient space formalism and Krein space quantization, including quantum metric fluctuation. An axiomatic quantum Yang-Milles theory can be constructed by generalizing this procedure to the spinor and vector fields. We recall that in Krein space with quantum metric fluctuation, the axioms are: a) dS covariance, b) locality, c) existence of an indefinite sesquilinear form, and d) analyticity properties.


\section{Color confinement and mass gap} \label{coco}

Let us briefly recall the results of Kharzeev et al. \cite{khletu}. Then the relationship with our model in \cite{ta223} will be discussed. They considered the coupling between the gluon fields, $F_{\mu\nu}^a(X)$, with the conformal sector of the metric $h(X)$, $g_{\mu\nu}=e^{h(X)}\eta_{\mu\nu}$. $X^\mu$ are the intrinsic coordinate systems. Using a Legendre transformation from the scalar field $h(X)$, they presented the dilaton field $\chi(X)$. Then they calculated the effective potential $W(\chi,F)$ in the constant chromomagnetic field, which explains the confinement problem of QCD in the one-loop approximation. The effect of geometry on gluon fields appears as the potential well, and Kharzeev et al. interpreted it as a conformal bag model. This behavior is extracted from the conformal sector of geometry or a scalar field $\chi(X)$. In this case, the scalar field is part of the geometry and can be considered a connection field or gauge potential, a common point with our model.

They obtained the glueball mass in terms of the dilaton mass, which can explain the mass gap problem. They calculated the leading radiative correction to the gluon propagator, which was constant, and they concluded that the strong coupling freezes at distances larger than the glueball size. Two crucial parts of their model are the scalar field as a gauge potential and a constant part in the propagator, which appear precisely in our model. Although their construction describes color confinement and mass gap problems in a one-loop approximation, their model is not renormalizable and cannot be formulated in an axiomatic QFT framework. These two last problems can be solved in our model due to using the dS ambient space formalism, Krein space quantization, and quantum metric fluctuation.

In the previous paper, we obtained that the gluon propagator in dS spacetime, at the early universe limit, $H\longrightarrow $ immense value, is a linear function of the ambient space coordinates $x^{\alpha}$ \cite{taga22}. Such a behavior of the two-point function freezes the color force for a long distance, and one can explain the color confinement in the early universe. We have also shown that due to the interaction between the gluon fields and the ghost fields, which are the mmc scalar fields, mass terms appear for the gluon field in the one-loop approximation. In the early universe, this mass term and linear coordinates behavior of the propagator explain the mass gap. In this limit, the effect of the geometry on QCD is significant, and the color confinement and mass gap can be easily seen \cite{taga22}.

First, the relation between our model and Kharzeev et al. model is discussed, and then we prove the color confinement and mass gap in the current universe. We know well that the conformal sector of the spacetime metric becomes a dynamical degree of freedom due to the trace anomaly of quantum matter fields \cite{anmo,anmamo}. In the Landau gauge of the gravitational field, the conformal sector becomes the mmc scalar field \cite{ta09,enrota}:
\b \label{consec} \mathcal{K}_{\alpha\beta}=\frac{1}{4}\theta_{\alpha\beta}\Phi_{\mathrm{m}}\, ,\e
where $\theta_{\alpha\beta}$ plays the role of the dS metric in ambient space formalism and $\Phi_{\mathrm{m}}$ is mmc scalar field. In this gauge, the interaction between the gluon fields, ($K_\alpha^a \,, \; a=1,\cdots,8$), and the mmc scalar fields $\Phi_{\mathrm{m}}$ appears as the geometry effect. It can be naturally extracted from the coefficients of connection of spacetime geometry, $\Gamma_{\mu\nu}^\rho$, in the covariant derivative.

Nevertheless, we know that the QFT in curved spacetime suffers from renormalizability. Krein space quantization (including quantum metric fluctuation) must be used to solve this problem; see \cite{ta02} and the references therein. Now we use the idea from the paper \cite{khletu} for the interaction of the conformal sector of the gravitational field in dS spacetime with gluon fields and combine it with quantization in Kerin space, which can explain the color confinement and mass gap in a renormalized covariant way.

We recall the Yang-Mills theory in dS ambient space formalism. The SU$(3)$ gauge invariant Lagrangian density for the gauge vector field and spinor field in the simplest form can be written in the following form \cite{taga22}:
\b \label{vespgain} \mathcal{ L}(K^a,\psi, \psi^\dag)=-\frac{1}{4}F_{\alpha\beta}^{\;\;\;\;a}F^{\alpha\beta a}+H\psi^\dag \gamma^0\left(-\ii\slashed{x}\gamma\cdot D^{K}+2\ii \right)\psi\, ,\e
where
\b \label{vespgain2} F_{\alpha\beta}^{\;\;\;\;a}=\nabla^\top_\alpha K_\beta^{\;\;a}-\nabla^\top_\beta K_\alpha^{\;\;a}+ g'C_{bc}^{\;\;\;\;a}K_\alpha^{\;\;b}K_\beta^{\;\;c}\,,\;\;\; \;\; x^\alpha F_{\alpha\beta}^{\;\;\;\;a}=0= x^\beta F_{\alpha\beta}^{\;\;\;\;a}\,,\e
and
\b D_\beta^K \psi \equiv \left(\nabla^\top_\beta -\ii g' K_{\beta}^{a}t_a\right)\psi\, .\e
$g'$ is the coupling constant between the spinor field and the gluon fields as the gauge potentials. From now on, and for simplicity, we ignore the index $a$.

If we consider the mmc scalar field as a gauge potential \cite{ta223}, it may be considered as a conformal part of the metric or the gravitational field. In this regard, it must appear in the connection coefficients of the geometry of spacetime,
$\Gamma_{\mu\nu}^\rho(\Phi_m\,, \cdots)$. In this case, spacetime geometry is no more Riemannian geometry and may be identified as Weyl geometry. For the definition of the covariant derivative of a vector field in Weyl geometry, see the equation $(37)$ in \cite{wh}. In the first-order perturbation on the dS covariant derivative, similar to the scalar-spinor field interaction \cite{ta223} and Weyl geometry, the covariant derivative of the vector field must be replaced with the following gauge-covariant derivative:
\b \label{tcdv} \nabla^\top_\alpha K_\beta \;\; \Longrightarrow \;\; D_{\alpha}^\Phi K_\beta \equiv \left(\nabla^\top_\alpha+g B^\top_\alpha \Phi_{\mathrm{m}}+\cdots \right)K_\beta\,,\e
where $g$ is the coupling constant between the gauge scalar field and the vector field $K_\beta$. $B^\alpha$ is an arbitrary constant five-vector field. $\nabla^\top_\alpha K_\beta=\partial^\top_\alpha K_\beta- H^2 x_\beta K_\alpha \,$ is the transverse derivative of a vector field on dS hyperboloid. For more discussion about the constant vector field $B^\alpha$, see \cite{ta223}.

By replacing the equation \eqref{tcdv} in the equations \eqref{vespgain} and \eqref{vespgain2}, the interaction between the scalar gauge field and the vector field can be extracted from the following Lagrangian density:
\b \label{intl} \mathcal{L}(K,\Phi_m)= -\frac{1}{2} \left(\nabla^\top_\alpha+ g B^\top_\alpha \Phi_m+\cdots \right)K_\beta \left(\nabla^{\top\alpha}+ g B^{\top\alpha} \Phi_m+\cdots \right)K^\beta+\cdots\,.\e
We can see that there appears a mass term for the vector field, which is a function of the mmc scalar field and the dS geometry:
\b \label{masterm} \mathcal{L}(K,\Phi_{\mathrm{m}})=\cdots +\frac{1}{2} g^2 B^\top \cdot B^\top \Phi_{\mathrm{m}}^2 \,K\cdot K +\cdots\,. \e
The Lagrangian density of the interaction field, including the symmetry breaking, results in a mass for gluon and then color confinement. Like the standard model, the vacuum expectation value of the scalar field, $\langle \Omega|\Phi_m|\Omega \rangle \equiv \mu\,$ plays the central role of the gauge symmetry breaking in this case. By replacing $\Phi_m$ with $\Phi_m+\mu\,$ in the Lagrangian density (\ref{masterm}) a mass generation occurs, for more details see \cite{ta223}. We obtain a mass for the gluon fields as
\b \label{massterm} m_g^2 \approx g^2\mu^2 B^\alpha B^\beta \theta_{\alpha\beta}\,.\e
This mass depends on the coupling constant $g$, the vacuum expectation value of the scalar field $\mu$, the constant vector field $B^\alpha$, and the dS spacetime geometry $\theta_{\alpha\beta}$. This mass in the gluon propagator can be interpreted as a representation of a short-length force

By comparing with the Kharzeev et al. paper \cite{khletu}, this mass term may be interpreted as the source of the spectral density in the Kall\'en-Lehmann type representation for the gluon propagator. As a result, the strong coupling freezes at long distances. The mixed gluon-scalar contributions to the spectral density appear only for the nonabelian gauge theory, which manifests color confinement in this case; for more details, see section III in \cite{khletu}.

We recall that a free field is called ``massive'' when it propagates inside the light cone and corresponds to a massive Poincar\'e representation in the null curvature limit. We call a free field ``massless'' if it propagates on the dS light cone and corresponds to a massless Poincar\'e representation at the null curvature limit. In this regard, the mmc scalar field is not a "true" massless field since it does not propagate only on the dS light cone, {\it i.e.} a constant term appears in the field propagator on the causal part of spacetime \cite{ta3}. The mass term \eqref{massterm} can also be associated with the constant solution of the mmc scalar field, and its propagation inside the light cone at the quantum field theory level \cite{ta223}.

In dS spacetime, the mmc scalar field quantization causes severe problems in the standard QFT approach \cite{allen85}. The problems can be overcome using Krein space quantization \cite{gareta00}. In this method, the dS invariant is preserved, and we have the propagation on and inside the light cone. Although this field is essential in QCD, linear quantum gravity, and quantum cosmology, it also plays a central role in quantum geometry when considered as the gauge potential \cite{ta223}.

In summary, we need the Krein space quantization and light-cone fluctuations (or quantum metric fluctuations) to describe the renormalisability of quantum fields in curved spacetime. Furthermore, for constructing an axiomatic quantum Yang-Mills theory, which includes color confinement and mass gap, we also need Krein space quantization, quantum metric fluctuations, and the conformal sector of quantum gravity or the mmc scalar field as the gauge potential. Therefore, there is a complicated entanglement between QFT and quantum geometry or gravity, which will be discussed in the following article.


\section{Conclusions}
\label{conclu}

In this article, we have supplemented our previous work regarding quantum Yang-Mills theory, {\it i.e.} color confinement in the current universe, and the analyticity properties of the vacuum expectation value of the product of the field operators. The two mathematical subjects, Krein space quantization, and dS ambient space formalism permit us to construct an axiomatic QFT for the Yang-Mills theory. The color confinement and mass gap problems can be extracted from the interaction between the gluon fields and the mmc scalar field as the conformal sector of the gravitational field. The Krein space quantization plays a central role in our model. The mmc scalar field is a connection between gauge theory and quantum dS geometry. We now have the building blocks needed to consider quantum dS geometry, which will be discussed in the following article \cite{tafinal}.

\vskip 0.5 cm
\noindent {\bf{Acknowledgements}}: The author thanks Jean Pierre Gazeau and Eric Huguet for their discussions and would like to thank le Coll\`ege de France, l'Universit\'e Paris Cit\'e, and  Laboratoire APC for their financial support and hospitality.


\end{document}